\documentclass[%
 reprint,
 amsmath,amssymb,
 aps,
]{revtex4-1}

\usepackage{graphicx}
\usepackage{dcolumn}
\usepackage{bm}
\usepackage{color}

\newcommand{\simgt}{\lower.5ex\hbox{$\; \buildrel > \over \sim \;$}}
\newcommand{\simlt}{\lower.5ex\hbox{$\; \buildrel < \over \sim \;$}}
\newcommand{\ch}[1]{{\textcolor{black}{#1}}}

\begin{document}

\preprint{APS/123-QED}

\title{Perturbation theory for the redshift-space matter power spectra after reconstruction}

\author{Chiaki Hikage}
\affiliation{%
 Kavli Institute for the Physics and Mathematics of the Universe (Kavli IPMU, WPI), University of Tokyo, 5-1-5 Kashiwanoha, Kashiwa, Chiba, 277-8583, Japan
}%
 \email{chiaki.hikage@ipmu.jp}
\author{Kazuya Koyama}%
\affiliation{%
 Institute of Cosmology and Gravitation, University of Portsmouth, Portsmouth PO1 3FX, UK
}%
\author{Ryuichi Takahashi}%
\affiliation{%
 Faculty of Science and Technology, Hirosaki University, 3 Bunkyo-cho, Hirosaki, Aomori 036-8588, Japan
}%

\date{\today}

\begin{abstract}   
We derive the one-loop perturbative formula of the redshift-space
matter power spectrum after density field reconstruction in the Zeldovich approximation. We find that the reconstruction reduces the amplitudes of nonlinear one-loop perturbative terms significantly by partially erasing the nonlinear mode-coupling between density and velocity fields. In comparison with N-body simulations, we find that both the monopole and quadrupole spectra of reconstructed matter density fields agree with the one-loop perturbation theory up to higher wavenumber than those before reconstruction. We
also evaluate the impact on cosmic growth rate assuming the survey volume and the number density like the Baryon Oscillation Spectroscopic Survey and find that the total error, including statistical and systematic ones due to one-loop approximation, decreases by half.
\end{abstract}

\maketitle

\section{Introduction}
Large-scale structure in the Universe is a powerful cosmological
probe to understand the properties of dark matter and dark energy
\citep[e.g.,][]{Amendola05}. Baryonic Acoustic Oscillations (BAO)
imprinted on the large-scale structure plays a role as a standard
ruler 
\citep{Eisenstein98,Meiksin99,BlakeGlazebrook03,HuHaiman03,Matsubara04,Angulo05,SeoEisenstein05,White05,Eisenstein07a,Huff07,Angulo08}
to determine the expansion history of the Universe from various galaxy surveys
\citep{Eisenstein05,Cole05,Padmanabhan07,Percival07,Okumura08,Xu13,Anderson14,Tojeiro14,Kazin14,Ross15,Alam16,Beutler17}. The
overall shape of the matter power spectrum is useful to infer the
neutrino mass \citep{Takada06,Saito10}. The anisotropy in the redshift-space clustering 
due to the bulk motion of galaxies provides a key probe to test General Relativity
\citep[e.g.,][]{Guzzo08,Yamamoto08,Reid12,Beutler13,Samushia13,Oka13,HY2013}.
One can expect precision cosmological analysis from galaxy clustering in upcoming galaxy surveys
such as the Prime Focus Spectrograph (PFS) \citep{PFS}, the Dark Energy Spectroscopic Instrument (DESI)
\citep{DESI16}, the Hobby-Eberly Telescope Dark Energy Experiment (HETDEX) \citep{HETDEX}, Euclid \citep{Euclid16}, the Wide Field Infrared Survey Telescope (WFIRST)
\citep{WFIRST15}.  

Nonlinearity in the gravitational evolution of large-scale structure makes precise cosmological analysis complicated. The BAO feature is degraded with
structure formation mainly due to the bulk motions of matter
\citep[e.g.,][]{CrocceScoccimarro08}. The perturbation theory has been
derived to describe the nonlinear effects on the power spectra
\citep[e.g.,][]{Vishniac83,Fry84,Goroff86,SutoSasaki91,Makino92,JainBertschinger94,Bouchet95,ScoccimarroFrieman96,Bernardeau02,JeongKomatsu06},
however, the availability is limited to the weakly nonlinear regime even
including the higher-order nonlinear terms
\citep{Matsubara08a,Matsubara08b,TNS10,Nishimichi11,Taruya12,Valageas13}.
Evolved density fields no longer follow Gaussian statistics and thereby their clustering information is not
fully described with two-point statistics but leaks to higher-order statistics.

\citet{Eisenstein07b} applies a density field reconstruction technique to aim for recovering
the original BAO signature by undoing the bulk motion in the 
Zeldovich approximation \citep{Zeldovich70}. The method has been extensively studied analytically and tested using numerical simulations
\citep{Seo08,Padmanabhan09,Noh09,Seo10,Sherwin12} and applied to the current BAO analysis \citep{Padmanabhan12,Xu13,Anderson14,Tojeiro14,Kazin14,Ross15,Alam16,Beutler17}.
It is also shown that the density field reconstruction recover the initial
density field out to smaller scales using
more optimal ways of density field reconstruction beyond the standard reconstruction method \citep{Seo10,TassevZaldarriaga12,Schmittfull15,Seo16,Schmittfull17,Wang17,Yu17,Hada18}.

Although the reconstruction succeeds in the BAO
analysis, it is relatively unclear how the reconstructed power spectrum can be
described in a perturbative manner.  In this paper, we derive the exact
one-loop order perturbative formula of the redshift-space matter power
spectra after reconstruction. In our previous work, we derive the one-loop perturbative formula of real-space matter power spectra and find that the amplitudes of the one-loop terms decrease significantly and then the perturbation theory can be applied to higher wavenumber $k$.
The result is consistent with the previous work showing that the reconstructed field better recovers the initial density field \citep[e.g.,][]{Seo10}. In this paper, we extend our previous work to redshift-space matter density fields. Recently \citet{Chen19} presented the
perturbative formula of halo power spectra in redshift space. Our
analysis is limited to the matter power spectra in redshift space, but include the nonlinearities from the Lagrangian to Eulerian mapping in our perturbative formula. From the comparison with large-scale suite of N-body simulations, we study to what extent the monopole and
quadrupole spectra of the redshift-space matter fields can be described in one-loop order. We also demonstrate
the impact on the cosmic growth rate assuming the survey volume and the number density like the Baryon Oscillation Spectroscopic Survey (BOSS) \citep{Alam16} when modeling the redshift-space power spectra with the one-loop perturbation.

The paper is organized as follows: Section \ref{sec:PT} describes the one-loop 
perturbation theory of the redshift-space matter power spectra and explicitly show the one-loop results in the Appendix \ref{sec:app}. In Section \ref{sec:results}, 
we study how the one-loop perturbation better describes the redshift-space matter power spectra in comparison with N-body simulations. Section \ref{sec:summary} is devoted to 
summary and conclusions.

\section{One-loop standard perturbation theory of redshift-space matter power spectra}
\label{sec:PT}
In this section, we derive the perturbative formula based on the standard perturbation theory (SPT)
to describe the nonlinearity in the redshift-space matter power
spectrum at one-loop order.

The comoving redshift space-position \ch{$\mathbf{x}$} is related to the
Lagrangian position $\mathbf{q}$ as
\begin{equation}
\mathbf{x}=\mathbf{q}+\mathbf{\Psi}^{z}(\mathbf{q}),
\end{equation}
where $\mathbf{\Psi}^{z}$ is the comoving displacement in
redshift space given by
\begin{equation}
\mathbf{\Psi}^{z}=\mathbf{\Psi}+\frac{\mathbf{\hat{z}\cdot\dot{\Psi}}}{H}\mathbf{\hat{z}},
\end{equation}
where $H$ is the time-dependent Hubble parameter and
$\mathbf{\hat{z}}$ is the unit vector of the line-of-sight direction.
\ch{In the Einstein-de-Sitter (EdS) model}, the $n$-th order perturbative displacement $\mathbf{\Psi}^{\rm
  (n)}$ is proportional to $n$-th power of the linear growth factor
$D$ and thereby the time derivative of the displacement becomes
\begin{equation}
\mathbf{\dot\Psi}^{\rm (n)}=nHf\mathbf{\Psi}^{\rm (n)},
\end{equation}
where $f=d\ln D/d\ln a$ is the linear growth rate. \ch{In a $\Lambda$CDM cosmology, the EdS approximation is valid to less than a percent level at the one-loop order of power spectra on the scales of our interest \citep{Bernardeau94,Fosalba98,Takahashi08,BoseKoyama16}}
The $n$-th order displacement in redshift space then becomes
\begin{equation}
\mathbf{\Psi}^{z(n)}=\mathbf{R}^{(n)}\mathbf{\Psi}^{(n)},
\end{equation}
where 
\begin{equation}
\mathbf{R}^{(n)}_{ij}=\delta_{ij}+nf\hat{z}_i\hat{z}_j,
\end{equation}
and $\delta_{ij}$ is Kronecker delta. The perturbative kernels in redshift space is given by
\begin{equation}
\mathbf{L}^{z(n)}=\mathbf{R}^{(n)}\mathbf{L}^{(n)}.
\end{equation}

The shift field $\mathbf{s}_z\mathbf{(x)}$ in redshift space is computed from the
negative ZA \citep{Zeldovich70} of the smoothed density field as
\begin{eqnarray}
\label{eq:sx}
\mathbf{s}^z\mathbf{(x)}&=&\int \frac{\mathbf{dk}}{(2\pi)^3}~\mathbf{\tilde{s}_k}^z~e^{i\mathbf{k\cdot x}}, \\
\label{eq:sk_ZA}
\mathbf{\tilde{s}_k}^{z}&=&-iW(k)\mathbf{L}^{(1)}(\mathbf{k})\tilde\delta_\mathbf{k}^{z},
\end{eqnarray}
where $W(k)$ is the smoothing kernel and we adopt a Gaussian kernel
$W(k)=\exp(-k^2R_s^2/2)$ with the smoothing scale of $R_s$. We found that the perturbation works best around $R_s=10h^{-1}$Mpc in real space\citep{HKH17} and thereby we fix $R_s$ to be 10$h^{-1}$Mpc in this paper.
The perturbative series of the shift field is given by
\begin{eqnarray}
\label{eq:skn_ZA}
\mathbf{\tilde{s}_k}^{z(n)}=-iW(k)\mathbf{L}^{(1)}(\mathbf{k})\tilde\delta_\mathbf{k}^{z(n)}.
\end{eqnarray}
where $\delta_\mathbf{k}^{z(n)}$ is the $n$-th order perturbation of
the redshift-space density fluctuation. 
This can be rewritten as
\begin{eqnarray}
\mathbf{\tilde{s}_k}^{z(n)}&=&\frac{iD^n}{n!}
\int\frac{\mathbf{dk}_1\cdot\cdot\cdot \mathbf{dk}_n}{(2\pi)^{3n-3}} 
\delta_{\rm D}\left(\sum_{j=1}^n \mathbf{k}_j-\mathbf{k}\right) 
\nonumber \\
&&\times
\mathbf{S}^{z(n)}(\mathbf{k}_1,...,\mathbf{k}_n)
\tilde\delta^{\rm L}_{\mathbf{k}_1}\cdot\cdot\cdot\tilde\delta^{\rm L}_{\mathbf{k}_n},
\end{eqnarray}
where the kernel of the shift field $\mathbf{S}^{z(n)}$ is written with the redshift-space Eulerian kernel $F_n^{z}$ as
\begin{equation}
\label{eq:shiftkernel}
\mathbf{S}^{z (n)}(\mathbf{k}_1,...,\mathbf{k}_n)=
-n!W(k)\mathbf{L}^{(1)}(\mathbf{k})F_n^{z}(\mathbf{k}_1,...,\mathbf{k}_n).
\end{equation}
The redshift-space kernel is given in the previous literature
\citep[e.g.,][]{Heavens98,Scoccimarro99,Matsubara08a}: 
\begin{equation}
F_1^{z}=1+f\mu^2,
\end{equation}
\begin{eqnarray}
F_2^{z}(\mathbf{k}_1,\mathbf{k}_2)&=&F_2(\mathbf{k}_1,\mathbf{k}_2)+f\mu^2G_2(\mathbf{k}_1,\mathbf{k}_2) \nonumber \\
&& +\frac12 fk\mu\left(\frac{k_{1z}}{k_1^2}+\frac{k_{2z}}{k_2^2}\right)
+\frac12 (fk\mu)^2\frac{k_{1z}k_{2z}}{k_1^2k_2^2}, \nonumber \\
\end{eqnarray}
\begin{eqnarray}
F_3^{z}(\mathbf{k}_1,\mathbf{k}_2,\mathbf{k}_3)&=&
F_3(\mathbf{k}_1,\mathbf{k}_2,\mathbf{k}_3)+f\mu^2G_3(\mathbf{k}_1,\mathbf{k}_2,\mathbf{k}_3) \nonumber \\
&& + fk\mu\frac{k_{1z}}{k_1^2}F_2(\mathbf{k}_2,\mathbf{k}_3) \nonumber \\
&& + fk\mu\frac{k_{2z}+k_{3z}}{|\mathbf{k_2+k_3}|^2}G_2(\mathbf{k}_2,\mathbf{k}_3) \nonumber \\
&& + (fk\mu)^2\frac{k_{1z}(k_{2z}+k_{3z})}{k_1^2|\mathbf{k_2+k_3}|^2}G_2(\mathbf{k}_2,\mathbf{k}_3) \nonumber \\
&& + \frac12 (fk\mu)^2\frac{k_{1z}k_{2z}}{k_1^2k_2^2}+\frac16 (fk\mu)^3\frac{k_{1z}k_{2z}k_{3z}}{k_1^2k_2^2k_3^2}, \nonumber \\
\end{eqnarray}
where $\mu=\mathbf{k\cdot\hat{z}}/k$ and $G_n$ is the $n$-th kernel of peculiar velocity field.

The displaced density field is written as
\begin{eqnarray}
\label{eq:deltak_unrecon}
\tilde\delta_\mathbf{k}^{\rm z(d)}=\int\mathbf{dq}e^{-i\mathbf{k\cdot q}}(e^{-i\mathbf{k}\cdot[\mathbf{\Psi}^z(\mathbf{q})+\mathbf{s}^z(\mathbf{x})]}-1),
\end{eqnarray}
where the shift field of the evolved mass particles is evaluated at
the Eulerian positions $\mathbf{x}$. The difference of the shift field
between the Eulerian and Lagrangian positions is perturbatively
expanded in terms of $\mathbf\Psi$ as
\begin{eqnarray}
\mathbf{s}^z(\mathbf{x})&=&\int \frac{\mathbf{dk}}{(2\pi)^3}
~\mathbf{\tilde{s}_k}^z~e^{i\mathbf{k\cdot (q+\Psi^z(q))}}, \\
&=&\sum_{n=0}^\infty\int \frac{\mathbf{dk}}{(2\pi)^3}~\mathbf{\tilde{s}_k}^z~e^{i\mathbf{k\cdot q}} \left[\frac{1}{n!}(i\mathbf{k\cdot \Psi^z(q)})^n\right], \nonumber \\
\label{eq:sx_pb}
&=&\mathbf{s}^z\mathbf{(q)}+\mathbf{(\Psi^z(q)\cdot\nabla)s}^z(\mathbf{q})
+\frac{1}{2}(\mathbf{\Psi}^z(\mathbf{q})\cdot\nabla)^2\mathbf{s}^z\mathbf{(q)}
\cdot\cdot\cdot . \nonumber \\
\end{eqnarray}
The shifted density field of a spatially uniform grid or random
is given by
\begin{equation}
\tilde\delta_\mathbf{k}^{z (s)}
=\int\mathbf{dq}e^{-i\mathbf{k\cdot q}}(e^{-i\mathbf{k\cdot s}^z(\mathbf{q})}-1),
\end{equation}
where the shift field of the (unevolved) uniform grid is evaluated at
the Lagrangian position. The reconstructed density field in redshift space is given as
\begin{eqnarray}
\label{eq:reconfield}
\tilde\delta_\mathbf{k}^{\rm z (rec)}&\equiv &
\tilde\delta_\mathbf{k}^{\rm z (d)} - \tilde\delta_\mathbf{k}^{\rm z(s)} \nonumber \\
&=&\int \mathbf{dq}e^{-i\mathbf{k\cdot q}}
e^{-i\mathbf{k\cdot s}^z(\mathbf{q})}(e^{-i\mathbf{k}
\cdot[\mathbf{\Psi}^z(\mathbf{q})+\mathbf{s}^z(\mathbf{x})
-\mathbf{s}^z(\mathbf{q})]}-1). \nonumber \\
\end{eqnarray}

The redshift-space formula is the same as that in real space but
replacing the real-space kernels $\mathbf{L}^{(n)}$ and $F_n$
with the redshift-space ones $\mathbf{L}^{z(n)}$ and $F_n^z$.

At linear order, the reconstructed density field in redshift space is not changed by reconstruction
\begin{eqnarray}
\label{eq:delrec1}
\delta_\mathbf{k}^{\rm z(rec) (1)}&=&\delta_\mathbf{k}^{z(1)}.
\end{eqnarray}
Higher-order terms of $\delta^{\rm (rec)}$ are given by
\begin{eqnarray}
\label{eq:F_n}
\tilde\delta_\mathbf{k}^{\rm z(rec,n)}&=&D^n(z)
\int\frac{\mathbf{dk}_1\cdot\cdot\cdot \mathbf{dk}_n}{(2\pi)^{3n-3}}\delta_{\rm D}
\left(\sum_{j=1}^n \mathbf{k}_j-\mathbf{k}\right) \nonumber \\
&&\times F_n^{\rm z (rec)}(\mathbf{k}_1,...,\mathbf{k}_n)
\tilde\delta^{\rm L}_{\mathbf{k}_1}\cdot\cdot\cdot\tilde\delta^{\rm L}_{\mathbf{k}_n}
\end{eqnarray}
where $F_n^{\rm z (rec)}$ is the Eulerian kernel for the reconstructed
matter density field in redshift space. 
We have already derived the explicit form of the reconstructed Eulerian kernel in real space in the previous paper \citep{HKH17}.  The first-order Eulerian kernel does not change after reconstruction
\begin{equation}
F_1^{\rm z(rec)}=F_1^{\rm z}.
\end{equation}
The second-order Eulerian kernel for the reconstructed field $F_2^{\rm z(rec)}$ can be derived 
by replacing the real-space kernel to the redshift-space one in the equation of (32) or (A10) in \citet{HKH17} as
\begin{eqnarray}
F_2^{\rm z(rec)}(&\mathbf{k}_1&,\mathbf{k}_2)=F_2^z(\mathbf{k}_1,\mathbf{k}_2) \nonumber \\
&&+\frac12\left[(\mathbf{k}\cdot \mathbf{S}^{\rm z(1)}(\mathbf{k}_1))
(\mathbf{k}_2\cdot \mathbf{L}^{z(1)}(\mathbf{k}_2))\right. \nonumber \\
&& +\left.(\mathbf{k}\cdot \mathbf{S}^{\rm z(1)}(\mathbf{k}_2))
(\mathbf{k}_1\cdot \mathbf{L}^{z(1)}(\mathbf{k}_1))\right].
\label{eq:F2_recon}
\end{eqnarray}
Note that $\mathbf{k}_i\cdot\mathbf{L}^{(1)}(\mathbf{k}_i)$ terms in the real
space becomes unity and thereby they are not written explicitly in the real-space formula.  In redshift space, however,
$\mathbf{k}_i\cdot\mathbf{L}^{(1)}(\mathbf{k}_i)$ becomes $1+f\mu_i^2$
where $\mu_i=\mathbf{k_i\cdot\hat{z}}$ and thus the $F_2^{\rm
  z(rec)}(\mathbf{k}_1,\mathbf{k}_2)$ depends on the line-of-sight
direction of the two wavenumbers $\mathbf{k}_1$ and
$\mathbf{k}_2$. This makes the one-loop perturbative formula complicated
as shown in Appendix \ref{sec:app}.

The third-order kernel is also derived by replacing the real-space kernel with the redshift-space one
in the equation of (33) or (A29) in \citep{HKH17} as
\begin{eqnarray}
F_3^{\rm z(rec)}(&\mathbf{k}_1&,\mathbf{k}_2,\mathbf{k}_3)=
F_3^z(\mathbf{k}_1,\mathbf{k}_2,\mathbf{k}_3) \nonumber \\
&&+\frac16\left[2(\mathbf{k\cdot S}^{z(1)}(\mathbf{k}_1))F_2^z(\mathbf{k}_2,\mathbf{k}_3)\right.
\nonumber \\
&&+(\mathbf{k\cdot S}^{z(1)}(\mathbf{k}_1))(\mathbf{k\cdot S}^{z(1)}(\mathbf{k}_2))(\mathbf{k_3}\cdot\mathbf{L}^{z(1)}(\mathbf{k_3}))
\nonumber \\
&& +(\mathbf{k\cdot S}^{z(2)}(\mathbf{k}_1,\mathbf{k}_2))
(\mathbf{k_3}\cdot\mathbf{L}^{z(1)}(\mathbf{k_3}))
\nonumber \\
&&+\left.{\rm (2~perms.)}
\right].
\label{eq:F3_recon}
\end{eqnarray}
The third-order kernel also
depends on the line-of-sight direction of three wavenumbers $\mu_i$
with $i=$1, 2, and 3.

The reconstructed power spectrum at one-loop order is written by 
\begin{equation}
P^{\rm z(rec),1-loop}(k)=D^2(z)P_{11}^{\rm z(rec)}(k)+D^4(z)(P_{22}^{\rm z(rec)}+P_{13}^{\rm z(rec)}),
\end{equation}
where $P_{nm}^{\rm z(rec)}=\langle\tilde\delta_\mathbf{k}^{\rm (n)}\delta_\mathbf{k}^{\rm (m)}\rangle$. The leading-order term is unchanged after reconstruction
\begin{eqnarray}
P_{11}^{\rm z(rec)}(k,\mu)=(1+f\mu^2)^2P_{\rm L}(k).
\end{eqnarray}
The one-loop terms of the redshift-space power spectrum can be 
written with the reconstructed Eulerian kernels as 
\begin{equation}
\label{eq:P22}
P_{22}^{\rm z(rec)}(k,\mu)=2\int \frac{\mathbf{dp}}{(2\pi)^3}
P_{\rm L}(|\mathbf{k-p}|)P_{\rm L}(p)[F_2^{\rm z(rec)}(\mathbf{k-p},\mathbf{p})]^2,
\end{equation}
and 
\begin{eqnarray}
\label{eq:P13}
P_{13}^{\rm z(rec)}(k,\mu)&=&6\ch{F_1^{\rm z(rec)}(k)}P_{\rm L}(k)\int \frac{\mathbf{dp}}{(2\pi)^3}P_{\rm L}(p)F_3^{\rm z(rec)}(\mathbf{k},\mathbf{p},\mathbf{-p}).
\nonumber \\
\end{eqnarray}
The exact formula of the one-loop terms are summarized in Appendix
\ref{sec:app}. The multipole components of the redshift-space power
spectrum is generally obtained by the Legendre polynomial expansion as
\begin{equation}
P_\ell(k)=\frac12\int_{-1}^1 d\mu P(k,\mu) {\cal L}_\ell(\mu),
\end{equation}
where ${\cal L}_\ell(\mu)$ is the Legendre polynomials, for example,
${\cal L}_0(\mu)=1$ and ${\cal L}_2(\mu)=(3\mu^2-1)/2$.

\begin{figure}
\begin{center}
\includegraphics[width=8.5cm]{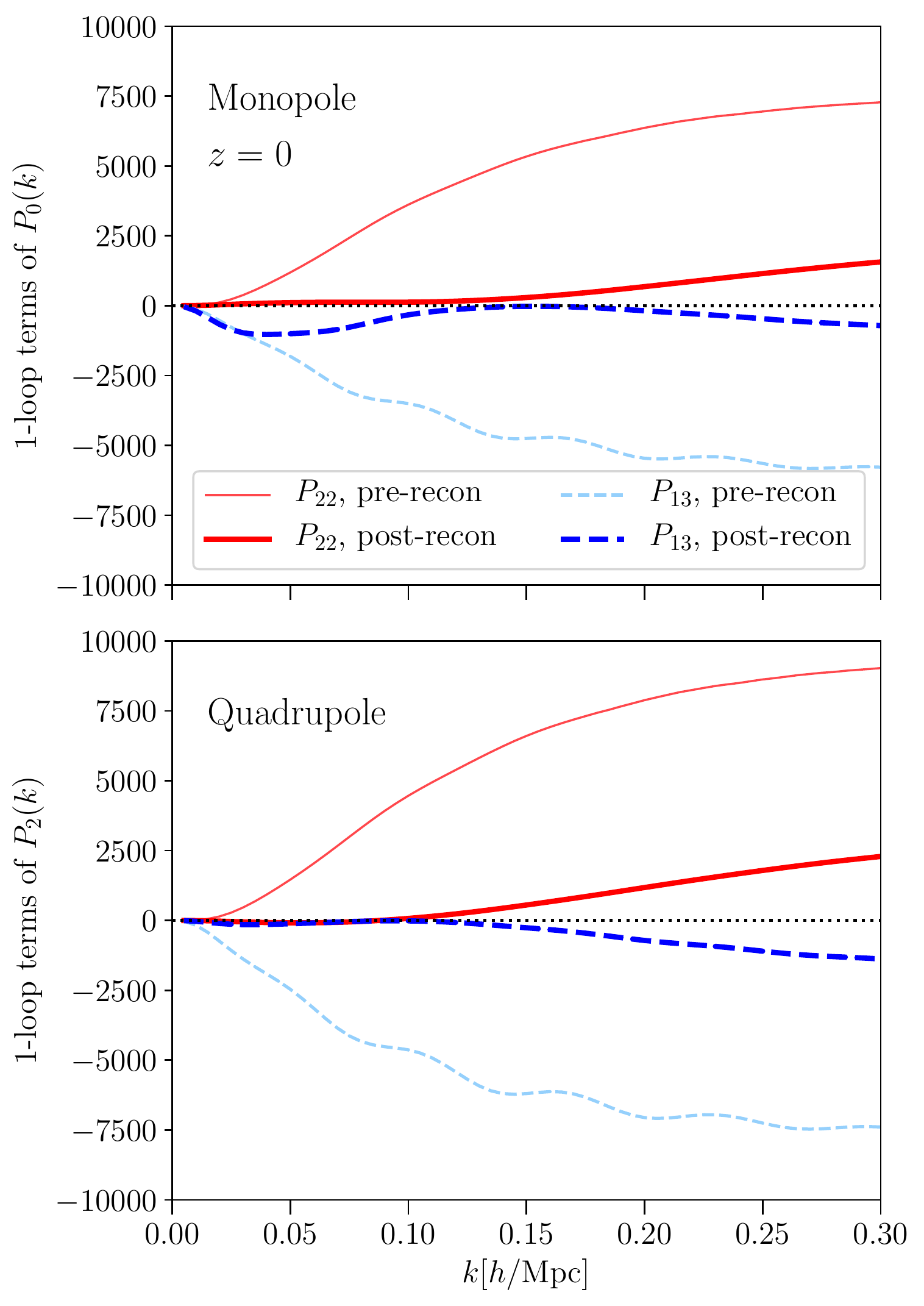}
\caption{Comparison of the one-loop terms $P_{22}(k)$ and $P_{13}(k)$ in monopole (upper) and quadrupole (lower) spectra of redshift-space matter density fields before (thin) and after (thick)
  reconstruction. The amplitudes of the both one-loop terms significantly decrease by reconstruction out to large $k$.}
\label{fig:stdPT}
\end{center}
\end{figure}

Figure \ref{fig:stdPT} shows the $k$-dependence of the one-loop terms $P_{13}$ and $P_{22}$ in the monopole ($\ell=0$) and quadrupole
($\ell=2$) spectra before and after reconstruction. 
We find that the amplitudes of both one-loop terms significantly decrease after
reconstruction in monopole and quadrupole spectra out to large $k$. This result is similar to the results in real space \citep[cf. Fig. 1 of ][]{HKH17}, but indicates that mode-couplings between density and velocity fields due to nonlinear gravity are 
partially removed by reconstruction. In-phase baryonic acoustic oscillations of $P_{13}$, but with negative amplitude, causes the degradation of the BAO signature. The oscillation of $P_{13}$ also significantly reduces after reconstruction and thereby the original BAO signature is substantially recovered. The result is consistent with that the BAO feature in redshift-space spectra is actually recovered by the reconstruction \citep[e.g.,][]{Padmanabhan12}.

\section{Results}
\label{sec:results}

\begin{figure*}
\begin{center}
\includegraphics[width=8.5cm]{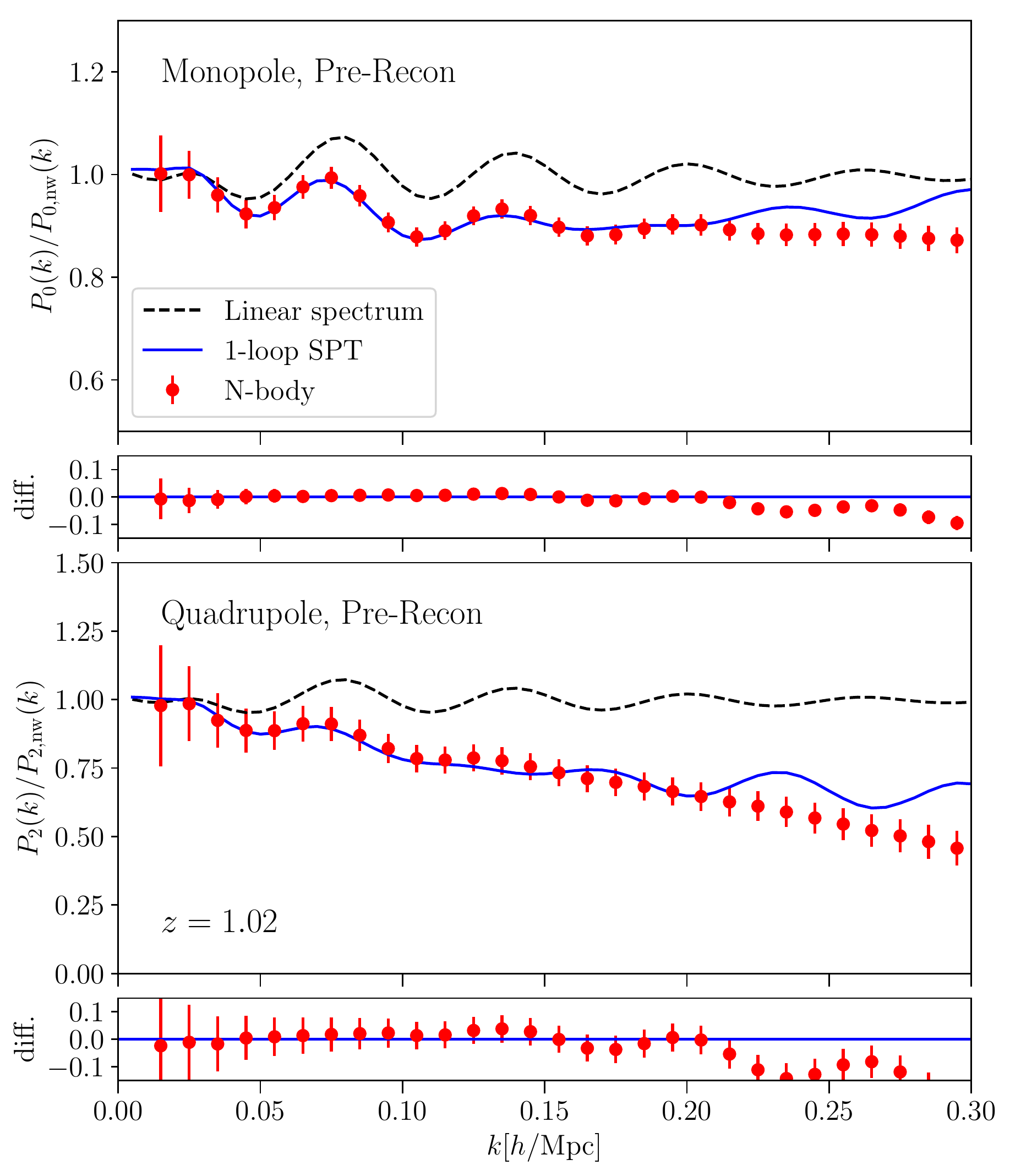}
\includegraphics[width=8.5cm]{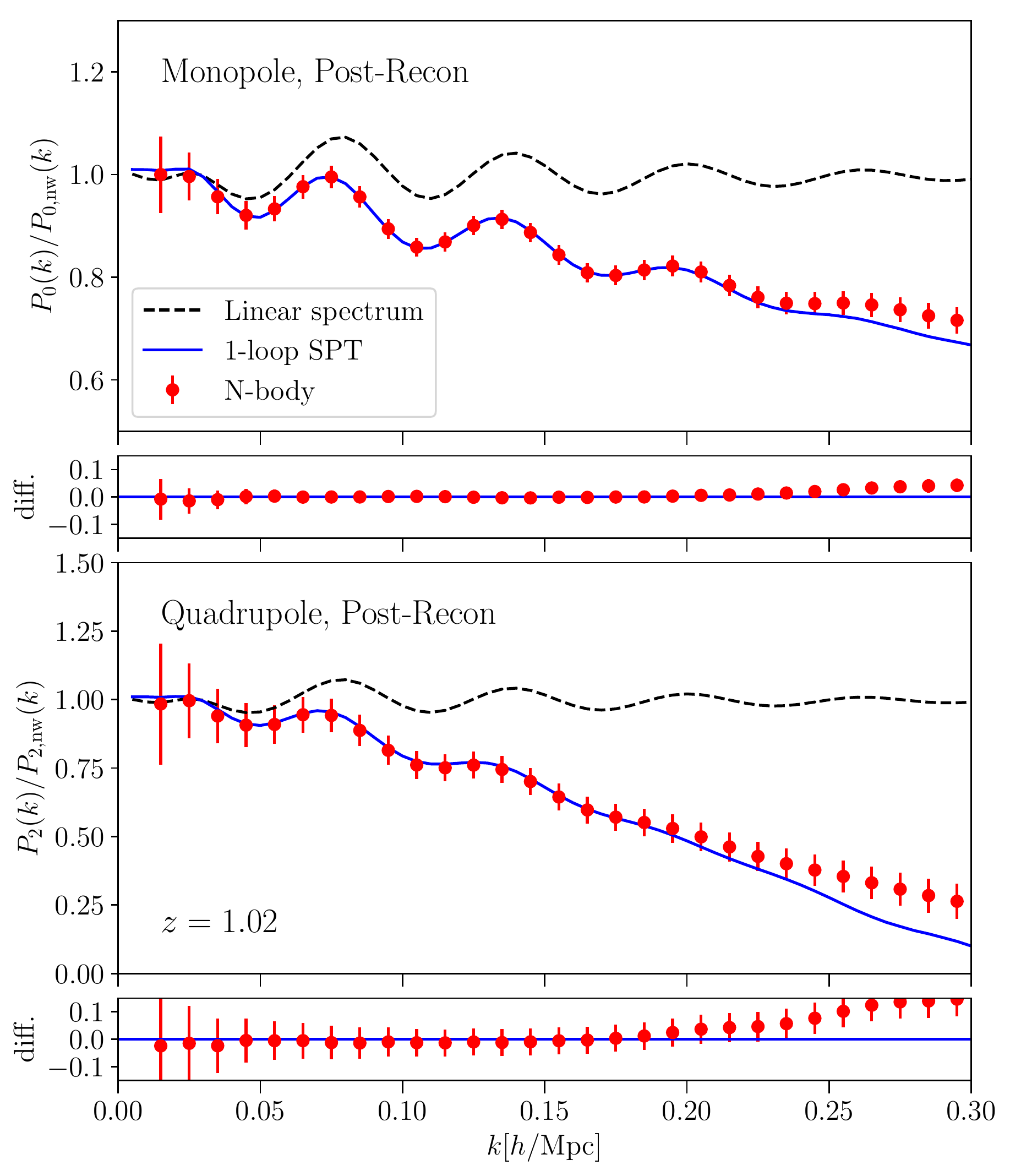}
\includegraphics[width=8.5cm]{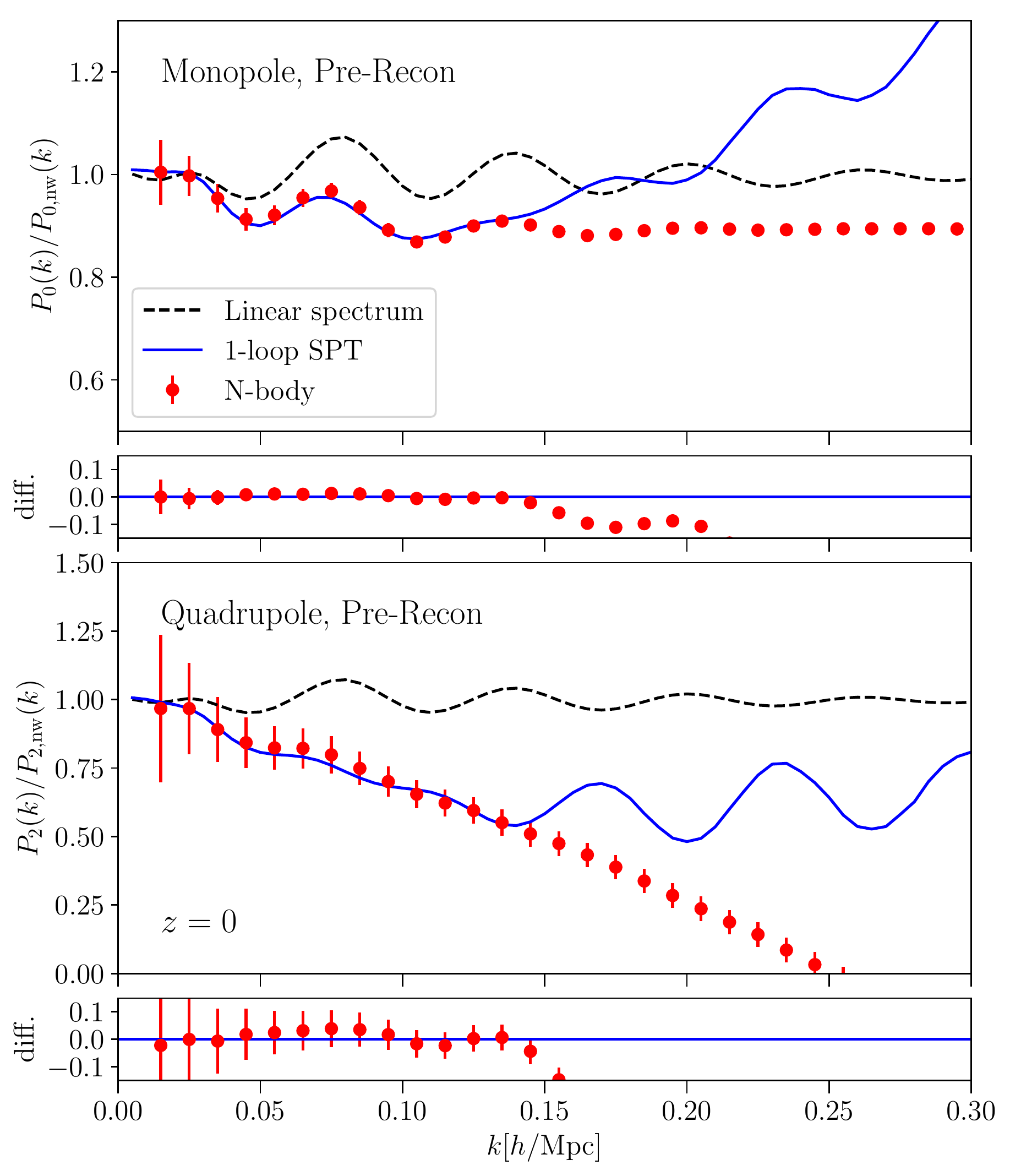}
\includegraphics[width=8.5cm]{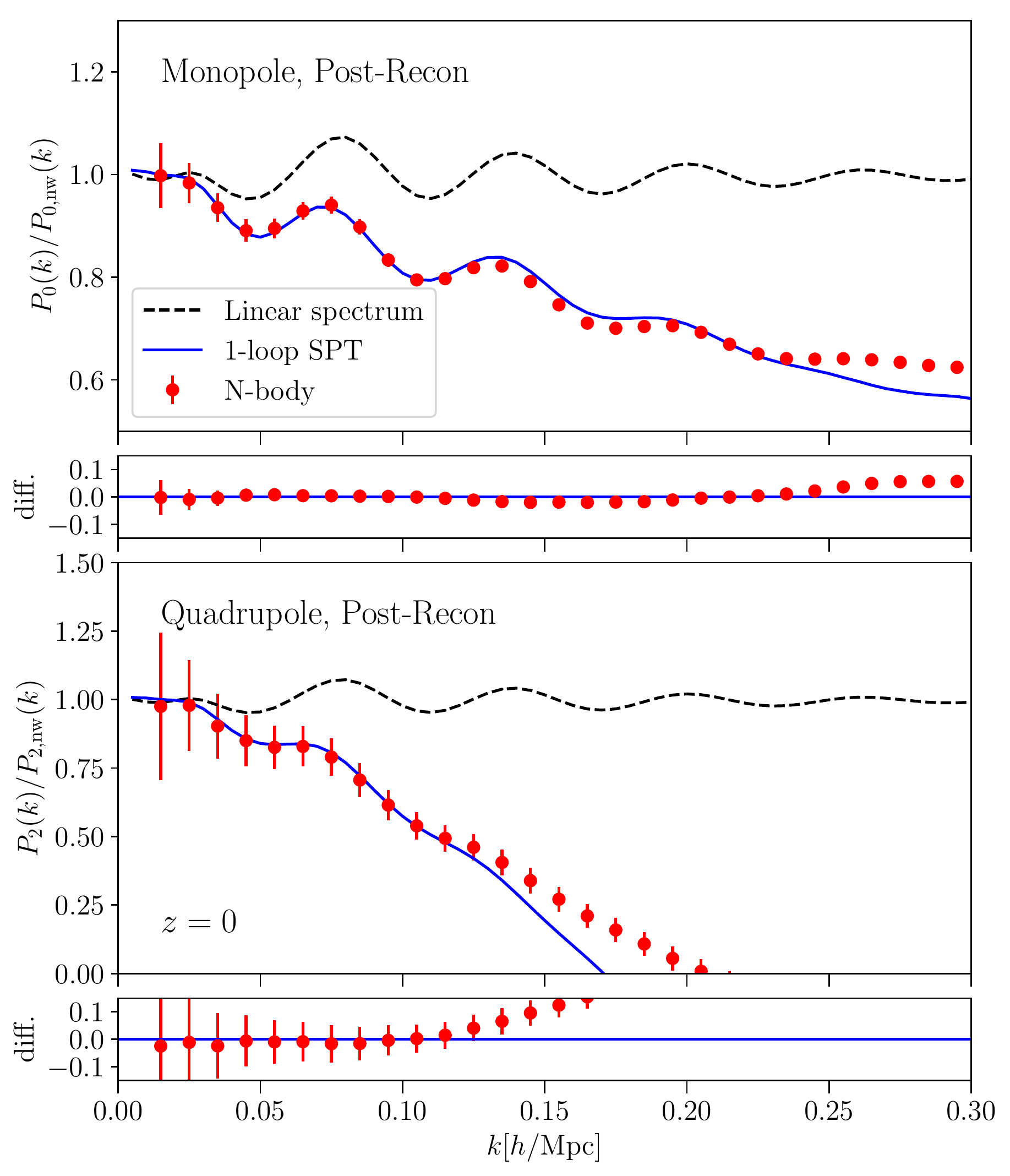}
\caption{Monopoles and quadrupoles of
  simulated redshift-space matter power spectrum (filled circles) before (left) and after (right) reconstruction at $z=1.02$ (upper) and $z=0$
  (lower). For comparison, we plot the one-loop SPT (solid lines) with two counterterms proportional to $P_\ell^{\rm (lin)}(k)k^2$ fitted to
  the simulated spectra upto $k=0.2h$/Mpc for $z=1.02$ and
  $k=0.12h$/Mpc for $z=0$.  For references, the linear spectra are plotted with lines. All of the power spectra are normalized with the no-wiggle
  multipole spectra in redshift space.  The plotted error-bars are Gaussian error assuming a BOSS-like survey with the survey volume $V=6
  (h^{-1}{\rm Gpc})^3$ and the number density $n=2\times 10^{-4}(h^{-1}{\rm Mpc})^{-3}$. Difference ratios between simulated spectra and one-loop spectra are also shown under each panel. The figure shows that their agreement becomes better after reconstruction and the perturbation works at higher $k$.}
\label{fig:pkl_sim}
\end{center}
\end{figure*}

\begin{figure}
\begin{center}
\includegraphics[width=8.5cm]{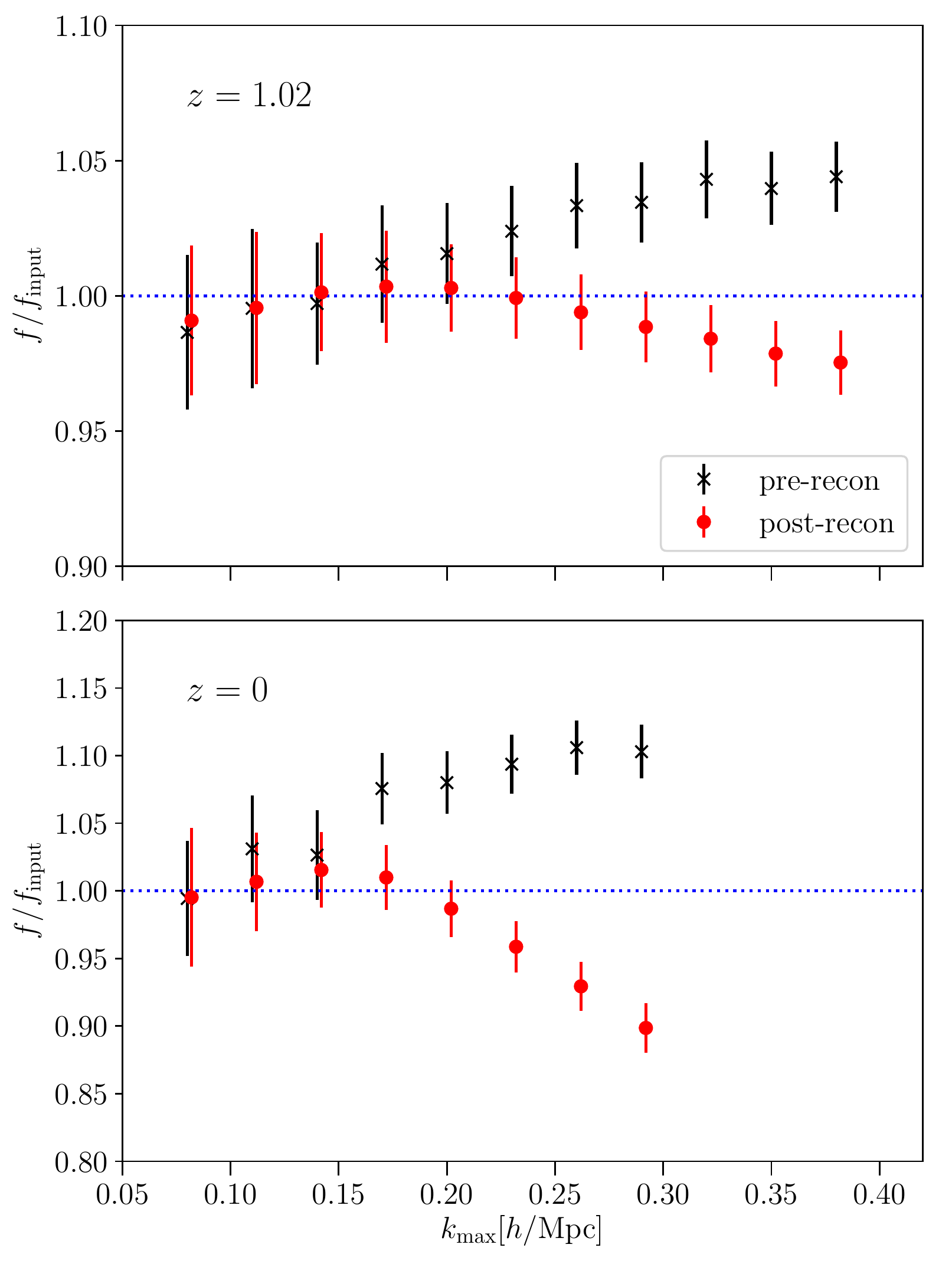}
\caption{Comparison of the $1\sigma$ statistical error and the systematic
  bias on the growth rate $f$ expected from the monopole and quadrupole matter
  power spectra as a function of $k_{\rm max}$ before and after 
  reconstruction. We again assume a BOSS-like survey volume and number density, i.e., $V=6
  (h^{-1}{\rm Gpc})^3$ and  $n=2\times 10^{-4}(h^{-1}{\rm Mpc})^{-3}$, to compute the error of the multipole power spectra, but the output redshift is $z=1.02$ (upper) and $z=0$ (lower) respectively. The one-loop perturbation
  theory is used to fit the simulated matter power spectra and thereby the
  systematic error becomes significant at higher $k_{\rm max}$. The figure shows
  that the reconstructed spectra better reproduces the input
  value of $f$ and then the systematic error exceeds the statistical
  error at higher $k_{\rm max}$ by reconstruction.}
\label{fig:fz}
\end{center}
\end{figure}

We compute the redshift-space matter power spectra using $N$-body simulations to see how well the one-loop perturbative formula describes the reconstructed spectra. Dark-matter $N$-body simulations are performed using a publicly available code {\tt Gadget-2} \citep{Springel05}. The mass particles are initially distributed based on 2LPT code \citep{CPS2006,Nishimichi2009} with Gaussian initial conditions at the input redshift of $31$. The initial linear power spectrum is computed by CAMB \citep{Lewis2000}. Each simulation is performed in a cubic box with the side length of $4h^{-1}{\rm Gpc}$ with $4096^3$ particles. 
We assign the $N$-body particles to $2048^3$ grid cells to calculate the density contrast, and then perform the Fourier transform \footnote{FFTW3 at http://www.fftw.org} to measure the power spectrum.
In our analysis, we use 8 realizations with two output redshifts of $z=0$ and $z=1.02$. 
We will show the average power spectrum with $1 \sigma$ error estimated from these realizations.
The cosmology in the simulations is based on a flat $\Lambda$CDM model with the best-fit values of Planck TT,TE,EE+lowP in 2015, i.e.,
$\Omega_b=0.0492$, $\Omega_m=0.3156$, $h=0.6727$, $n_s=0.9645$, and
$\sigma_8=0.831$ \citep{Planck15}. 

We evaluate the agreement between the simulated power spectra and the perturbative formula with the following $\chi^2$ value:
\begin{eqnarray}
\label{eq:chisq}
\chi^{2}&=&\sum_i^{k_{\rm min}\le k_i\le k_{\rm max}}\sum_{\ell\ell^\prime}^{0,2}[P_\ell^{\rm theory}(k_i)-P_\ell^{\rm sim}(k_i)] \nonumber \\
&& {\bf Cov}^{-1}_{\ell\ell^\prime}(k_i) [P_{\ell^\prime}^{\rm theory}(k_i)-P_{\ell^\prime}^{\rm sim}(k_i)],
\end{eqnarray}
where ${\bf Cov}_{\ell\ell^\prime}(k_i)$ represents the covariance of
multipole power spectra at a given $k_i$. Here we focus on the monopole and quadrupole components of redshift-space matter power spectra. For simplicity, we adopt the analytical formula of the Gaussian covariance given
by the Appendix C of \citet{TNS10} with typos corrected and neglect
the off-diagonal components of the covariance between different bins of $k$. This approximation would be
valid when the cosmic variance and/or shot-noise terms are dominant
compared to the non-Gaussian terms. The Gaussian covariance depends on
the survey volume $V$ and the number density $n$.  In this paper, we assume BOSS-like survey with $V=6(h^{-1}{\rm Gpc})^3$ and $n=2\times
10^{-4}(h^{-1}{\rm Mpc})^{-3}$.  The chi-squared value depends on the
range of $k$.  Here we fix the minimum value $k_{\rm min}=0.01h{\rm Mpc}^{-1}$
and see how $\chi^2$ changes as the maximum value $k_{\rm max}$ increases.
For the theoretical power spectrum $P^{\rm theory}(k)$, we adopt the one-loop perturbative formula with the lowest-order counter term proportional to $k^2 P_\ell$ in each $\ell$ 
\begin{equation}
P_\ell^{\rm theory}(k)=P_\ell^{\rm z,1-loop}(k)+\alpha_\ell k^2P_\ell^{\rm L}(k).
\end{equation}
The counter terms renormalize the contributions from UV
(small-scale) power \citep[e.g.,][]{eftoflss,Fonseca17} including the lowest-order contributions of nonlinear redshift-space
distortions, i.e., Fingers-of-God effect. The proportional factor
$\alpha_\ell$ ($\ell=0$ and $2$) are obtained by fitted them to the simulated power spectra. Note that we adopt one counter term per each multipole for both
pre-recon and post-recon spectrum, while \citep{Chen19} adopts three
counter terms per each multipole proportional to the power spectra for $\langle\tilde\delta_k^{z(d)}\tilde\delta_k^{\ast z(d)}\rangle$, $\langle\tilde\delta_k^{z(d)}\tilde\delta_k^{\ast z(s)}\rangle$,$\langle\tilde\delta_k^{z(s)}\tilde\delta_k^{\ast z(s)}\rangle$ for the reconstructed spectrum.

Figure \ref{fig:pkl_sim} shows the comparison of the monopole and quadrupole of simulated power
spectrum with the one-loop perturbative formulae before (left) and after reconstruction (right) at the output redshift of 1.02 (upper) and 0 (lower). We adopt the bestfit
values (the minimum $\chi^2$) of counter terms with $k_{\rm max}$ of 0.2$h$/Mpc for $z=1$ and 0.12$h$/Mpc for $z=0$ where the minimum
$\chi^2$ value is less than unity. In these plots, the power spectra are normalized
with redshift-space no-wiggle spectrum including linear Kaiser effect
\citep{Kaiser87}, i,e., $P^{\rm nw,z}(k,\mu)=(1+f\mu^2)^2P^{\rm
  nw}(k)$ where $f\equiv d \ln D(z)/d \ln a$ is the growth rate at a
given $z$ and $P^{\rm nw}(k)$ is the no-wiggle spectrum given by
\citet{EisensteinHu98}.  We find that the one-loop perturbative
formula can be better fitted to the post-recon spectrum up to higher
$k$. More quantitatively saying, $k_{\rm max}$ where $\chi^2_{\rm min}$ becomes
unity is 0.17$h$/Mpc for $z=1$ and 0.11$h$/Mpc for $z=0$ before
reconstruction, which are extended to be 0.23$h$/Mpc for $z$=1 and
0.13$h$/Mpc for $z=0$ after reconstruction.

We also see the impact on the measurement of the growth rate $f$ by
computing the likelihood function ${\cal L}\propto \exp({-\chi^2/2})$
where $\chi^2$ is computed from the equation (\ref{eq:chisq}). In addition to the counter terms $\alpha_0$ and $\alpha_2$, the growth rate $f$ is treated as free parameters. Figure \ref{fig:fz} shows the expected constraints on $f$
from the monopole and quadrupole spectra with different
$k_{\rm max}$.  Here we again assume the error expected from the same BOSS-like
survey volume and number density. The statistical error 
decreases at higher $k_{\rm max}$, however, the systematic error increases because the
one-loop approximation becomes worse at higher $k$. We find that
$k_{\rm max}$ where the statistical error is comparable to the
systematic one is $0.22h$/Mpc for $z=1$ and $0.12h$/Mpc for $z=0$ before reconstruction. 
The corresponding wavenumbers are extended to be $0.30h$/Mpc for $z=1$ and $0.21h$/Mpc for $z=0$ after reconstruction. The errors at the $k_{\rm max}$ where the statistical
error is comparable to the systematic one decreases from 0.0171 to
0.0128 (40$\%$ decrement) for $z=1$ and from 0.0405 to 0.0197 (51$\%$ decrement) for $z=0$ by reconstruction.

\section{Summary and Conclusions}
\label{sec:summary}

We derived the one-loop perturbative formulae of the
redshift-space matter power spectra after density-field reconstruction using the Zeldovich approximation. We found that the amplitudes of the one-loop nonlinear terms $P_{13}(k)$ and
$P_{22}(k)$ decrease significantly in both monopole and quadrupole spectra. Our result indicates that the mode couplings among density and velocity fields associated with nonlinear gravity are partly eliminated by the reconstruction. From the comparison of N-body
simulations, we showed that the one-loop perturbative formulae better describe the monopole and quadrupole of matter power
spectra after reconstruction and agree with the simulated spectra at higher
$k$. We also estimated the impact on the measurement of the growth rate when
using the one-loop perturbation theory as a theoretical modeling of
the redshift-space matter power spectra assuming the survey volume and
number density of a BOSS-like galaxy survey. We found that the
systematics due to the one-loop approximation is reduced by
reconstruction and thereby the total error of the growth rate
measurement including the statistical and systematic errors decreases by half.

In this paper, we focused on the redshift-space matter power spectra. We plan to extend our analysis to the power spectra of biased tracers but leave this work in the near future.  In this analysis, we neglected the non-Gaussianity in the covariance of matter power spectra. Since the leading-order non-Gaussianity
also comes from the one-loop terms \citep{Mohammed17,Wadekar19}, the non-Gaussianity should be smaller after
reconstruction and thereby the information contents of the power spectrum is expected to increase by reconstruction.  We plan to show more detailed analysis of the covariance of reconstructed power spectra in the near future.

\begin{acknowledgments}
We thank Alan Heavens and Marcel Schmittfull for useful discussion.
This work is supported by MEXT/JSPS KAKENHI Grant Numbers
JP16K17684 (CH), JP18H04348 (CH), JP15H05893 (RT), and JP17H01131 (RT). KK is supported by the UK STFC grant ST/N000668/1 and ST/S000550/1, and the European Research Council under the European Union's Horizon 2020 programme (grant agreement No.646702 "CosTesGrav").
Numerical computations were in part carried out on Cray XC30 and XC50 at Centre for Computational Astrophysics, National Astronomical Observatory of Japan.

\end{acknowledgments}

\appendix

\begin{widetext}
\section{Derivation of the one-loop matter power spectrum in redshift space}
\label{sec:app}
The one-loop terms of the redshift-space matter power spectra $P_{13}(k)$ and $P_{22}(k)$ are
derived from the equations (\ref{eq:P22}) and (\ref{eq:P13}) after a
lengthy but straightforward calculation.  Their equations are
summarized below after the integration over the azimuthal angle of $\mathbf{p}$ as follows:
\begin{eqnarray}
P_{22}^{\rm z(rec)}(\mathbf{k})=
\sum_{n,m}\mu^{2n}f^{m}\frac{k^3}{4\pi^2} \int_0^{\infty}dr P_{\rm L}(kr)\int_{-1}^{1}dx P_{\rm L}(k(1+r^2-2rx)^{1/2}) \frac{A_{nm}(r,x)}{(1+r^2-2rx)^2},
\end{eqnarray}
and
\begin{eqnarray}
P_{13}^{\rm z(rec)}(\mathbf{k})=(1+f\mu^2)P_{\rm L}(k)\sum_{n,m}\mu^{2n}f^{m}\frac{k^3}{4\pi^2}
\int_0^{\infty}dr P_0(kr) \int_{-1}^{1}dx B_{nm}(r,x).
\end{eqnarray}
where $A_{nm}$ and $B_{nm}$ are the coefficients of $\mu^{2n}f^{m}$ terms in the one-loop terms.
The reconstructed spectra depends on the smoothing kernel in the equation (\ref{eq:sk_ZA}), which are used to derive the shift field from the smoothed density field, and thereby the following equations of the coefficients of the one-loop terms include $W(|\mathbf{p}|)$ and $W(|\mathbf{k-p}|)$ which are denoted as $W_p$ and $W_\star$ respectively. These equations agree with the SPT
calculation \citep{Matsubara08a} at
the limit of pre-reconstruction, i.e., $W_p\rightarrow 0$ and
$W_\star\rightarrow 0$.
The nonvanishing components of $A_{nm}$ and $B_{nm}$ are summarized below:
\begin{eqnarray}
A_{00}&=&\frac{\left(r \left(7 r x (W_\star-W_p)+2 (7 W_p-5) x^2-7 W_\star+3\right)-7 (W_p-1) x\right)^2}{98} \\
A_{01}&=&-\frac{1}{14 \left(r^2-2 r x+1\right)} 
  \left[\left(x^2-1\right) (2 r (r-x)+1) (r (W_p x (r-2 x)-r W_\star x+W_\star)+W_p x) \right. \nonumber \\
  && \times \left. \left(r \left(7 r x (W_p-W_\star)+2 (5-7 W_p) x^2+7 W_\star-3\right)+7 (W_p-1) x\right)\right] \\
A_{11}&=& \frac{1}{98 \left(r^2-2 r x+1\right)} 
   \left[\left(r \left(7 r x (W_\star-W_p)+2 (7 W_p-5) x^2-7 W_\star+3\right)-7 (W_p-1) x\right)\right. \nonumber \\
  && \left(-14 r^4 x \left(3 x^2-1\right) (W_p-W_\star)+2 r^3 \left(-21 x^4 (W_\star-3 W_p)-x^2 (7 W_p+28 W_\star+20)+7 W_\star+6\right) \right. \nonumber \\
  && +r^2 x \left(x^2 (-91 W_p+63 W_\star+80)-84 W_p x^4+7 W_p+21 W_\star+4\right) \nonumber \\
  && +r \left(x^2 (28 W_p-21 W_\star-96)+84 W_p x^4-7 W_\star+12\right) -\left.\left.7 x \left(3 W_p x^2+W_p-4\right)\right)\right] \\
A_{02}&=&\frac{3 \left(x^2-1\right)^2}{112 \left(r^2-2 r x+1\right)^2} 
    \left[\left(2 r^2 \left(r^2-2 r x+1\right) (-r (W_p x (r-2 x)-r W_\star x+W_\star)-W_p x) \right.\right. \nonumber \\
   && \left(r \left(7 r x (W_\star-W_p)+2 (7 W_p-5) x^2-7 W_\star+3\right)-7 (W_p-1) x\right) \nonumber \\
   && +\left.\left.7 (2 r (r-x)+1)^2 (r (W_p x (r-2 x)-r W_\star x+W_\star)+W_p x)^2\right)\right]
   \end{eqnarray}
   \begin{eqnarray}
A_{12}&=&-\frac{1}{56\left(r^2-2 r x+1\right)^2}\left[\left(x^2-1\right) \left(126 r^8 x^2 \left(5 x^2-1\right) (W_p-W_\star)^2\right.\right. \nonumber \\
   && +2 r^7 x (W_p-W_\star) \left(30 x^4 (-63 W_p+21 W_\star+5)+9 x^2 (14 W_p+84 W_\star+13)-126 W_\star-1\right) \nonumber \\
   && +2 r^6 \left(30 x^6 \left(133 W_p^2-2 W_p (49 W_\star+10)+W_\star (7 W_\star+10)\right)\right. \nonumber \\ 
  && +3 x^4 \left(497 W_p^2-W_p (1624 W_\star+295)+89 W_\star (7 W_\star+3)\right)\nonumber \\
   && +x^2 \left(112 W_p W_\star-W_p (119 W_p+111)+574 W_\star^2+230 W_\star+48\right) 
   \left.-63 W_\star^2-W_\star-6\right)+4 r^5 x \left(-105 W_p^2\right. \nonumber \\
  && \left(16 x^6+27 x^4+x^2\right) 
   +W_p \left(60 (7 W_\star+5) x^6+42 (84 W_\star+29) x^4+(1498 W_\star+501) x^2-56 W_\star+11\right) \nonumber \\
   && -\left.W_\star \left(35 W_\star \left(9 x^4+29 x^2+2\right)+642 x^4+655x^2+68\right)-96 x^2-9\right) \nonumber \\
   && +r^4 \left(x^4 \left(5691 W_p^2-14 W_p (881 W_\star+430)+15 W_\star (91 W_\star+270)+384\right) \right. \nonumber \\
   && +x^2 \left(-63 W_p^2-14 W_p (99 W_\star+56)+W_\star (1981 W_\star+1784)+480\right) \nonumber \\
   && \left.+1680 W_p^2 x^8+168 W_p x^6 (74 W_p-25 W_\star-27)+2 (W_\star (7W_\star+23)-5)\right) \nonumber \\
   && +2 r^3 x \left(-28 W_p^2 x^2 \left(60 x^4+158 x^2+21\right)+W_p \left(42 (45 W_\star+73) x^4+(2485 W_\star+1577) x^2+49 W_\star+47\right) \right. \nonumber \\
   && \left.-217 W_\star^2-(W_\star (315 W_\star+1493)+360) x^2-257 W_\star-88\right) \nonumber \\
   && +r^2 \left(x^2 \left(84 W_p^2-2 W_p (469 W_\star+351)+3 W_\star (35 W_\star+348)+488\right) \right. \nonumber \\
   && \left.+2520 W_p^2 x^6+2 W_p x^4 (1512 W_p-735 W_\star-1945)+35 W_\star^2+48 W_\star+16\right) \nonumber \\
   && +2 r x \left(W_p \left(x^2 (-252 W_p+105 W_\star+592)-420 W_p x^4+35 W_\star+24\right)-70 (W_\star+1)\right) \nonumber \\
   &&\left.\left.+35 W_p x^2 \left(3 W_p x^2+W_p-4\right)+14\right)\right] \\
A_{22}&=&\frac{1}{784 \left(r^2-2 r x+1\right)^2}\left[294 r^8 x^2 \left(35 x^4-30 x^2+3\right) (W_p-W_\star)^2\right. \nonumber \\
   &&+14 r^7 x (W_p-W_\star) \left(70 x^6 (-63 W_p+21 W_\star+5)+x^4 (2940 W_p+1050 W_\star+299) \right. \nonumber \\
   &&+\left.18 x^2 (7 W_p-91 W_\star-18)+126 W_\star+11\right) \nonumber \\
   && +2 r^6 \left(490 x^8 \left(133 W_p^2-2 W_p (49 W_\star+10)+W_\star (7 W_\star+10)\right) \right. \nonumber \\
   && +x^4 \left(-23520 W_p^2+14 W_p (3906 W_\star+673)-147 W_\star (45 W_\star+19)+2832\right) \nonumber \\
   && +x^2 \left(343 W_p^2+133 W_p (28 W_\star+13)-7 W_\star (1463W_\star+593)-1884\right) \nonumber \\
   && -7 x^6 \left((8540 W_p-2297) W_\star+5 W_p (259 W_p+509)-4375 W_\star^2\right) \nonumber \\
   && +\left. 441 W_\star^2+77 W_\star+228\right) \nonumber \\
   && +4 r^5 x \left(-7 x^6 \left(5215 W_p^2-2 W_p (4130 W_\star+1823)+W_\star (735 W_\star+1874)\right) \right. \nonumber \\
   && +x^4 \left(27146 W_p^2+7 W_p (151-294 W_\star)-49 W_\star (344 W_\star+239)-5664\right) \nonumber \\
   && +x^2 \left(3871 W_p^2-28 W_p (763 W_\star+233)+7 W_\star (1197 W_\star+850)+1878\right) \nonumber \\
   && -\left. 980 W_p x^8 (28 W_p-7 W_\star-5)-98 W_p W_\star-259 W_p+1862 W_\star^2+1239 W_\star+258\right) \nonumber \\
   && +r^4 \left(x^6 \left(8281 W_p^2-14 W_p (14679 W_\star+8494)+49 W_\star (455 W_\star+1942)+22656\right)\right. \nonumber \\
   && +2 x^4 \left(-36701 W_p^2+14 W_p (2989 W_\star+1318)+35 W_\star (490 W_\star+293)+13248\right) \nonumber \\
   && -7 x^2 \left(217 W_p^2-2 W_p (1841 W_\star+894)+W_\star (2891 W_\star+2914)+2084\right) \nonumber \\
   && +\left. 27440 W_p^2 x^{10}+56 W_p x^8 (3640 W_p-1225 W_\star-1699)-14 W_\star (77 W_\star+85)+716\right) \nonumber \\
   && +2 r^3 x \left(-392 W_p^2 \left(70 x^6+189 x^4-67 x^2-24\right) x^2 \right. \nonumber \\
   && +7 W_p \left(2 (2205 W_\star+5273) x^6+(6223 W_\star+3347) x^4-4 (847 W_\star+488) x^2-189 W_\star-181\right) \nonumber \\
   && -3 (245 W_\star (7 W_\star+53)+8816) x^4-14 W_\star (301 W_\star+13) x^2 \nonumber \\
   && +\left. 7 W_\star (329 W_\star+551)+592 x^2+2336\right) \nonumber \\
   && +r^2 \left(-14 W_\star \left(x^2 \left(7 W_p \left(245 x^4+190 x^2-99\right)-2124 x^2+12\right)+96\right) \right. \nonumber \\
   && +2 x^2 \left(7 W_p \left(2940 W_p x^6+(3836 W_p-7381) x^4-98 (20 W_p+3) x^2-112 W_p+619\right) \right. \nonumber \\
   && +\left.\left. 8\left(2881 x^2-680\right)\right)+49 W_\star^2 \left(35 x^4+18 x^2-5\right)+64\right) \nonumber \\
   && +14 r x \left(-980 W_p^2 x^6+W_p x^4 (-728 W_p+245 W_\star+2432)+2 x^2 (W_p (182 W_p+63 W_\star+8)-2 (77 W_\star+317)) \right. \nonumber \\
   && -\left.35 W_p W_\star-96 W_p-28 W_\star+260\right) \nonumber \\
   && +\left.49 \left(x^2 \left(W_p \left(35 W_p x^4+2 (9 W_p-44) x^2-5 W_p-8\right)+52\right)-4\right)\right]
   \end{eqnarray}
   \begin{eqnarray}
A_{03}&=&-\frac{5 r^2 \left(x^2-1\right)^3 (2 r (r-x)+1) (r (W_p x (r-2 x)-r W_\star x+W_\star)+W_p x)^2}{16
   \left(r^2-2 r x+1\right)^2} \\
A_{13}&=&\frac{1}{112 \left(r^2-2 r x+1\right)^2} \left[3 \left(x^2-1\right)^2 (r (W_p x (r-2 x)-r W_\star x+W_\star)+W_p x)\right. \nonumber \\
   && \left(70 r^6 x \left(7 x^2-1\right) (W_p-W_\star)+2 r^5 \left(245 x^4 (W_\star-3 W_p) \right.\right. \nonumber \\ 
   && +\left. x^2 (-105 W_p+420 W_\star+76)-35 W_\star+8\right) \nonumber \\
   && +r^4 x \left(x^2 (1715 W_p-1015 W_\star-304)+980 W_p x^4-7 W_p-413 W_\star-200\right) \nonumber \\
   && +r^3 \left(x^2 (-644 W_p+693 W_\star+488)-1540 W_p x^4+63 W_\star+58\right) \nonumber \\
   && +7 r^2 x \left(W_p \left(123 x^2+11\right)-26 W_\star-36\right) \nonumber \\
   && +\left. \left.14 r \left(-14 W_p x^2+W_\star+3\right)+14 W_p x\right)\right] \\
A_{23}&=&\frac{1}{112 \left(r^2-2 r x+1\right)^2} 
     \left[\left(x^2-1\right) \left(-210 r^8 x^2 \left(21 x^4-14 x^2+1\right) (W_p-W_\star)^2 \right.\right. \nonumber \\
   && +2 r^7 x (W_p-W_\star) \left(-2205 x^6 (W_\star-5 W_p)-10 x^4 (441 W_p+588 W_\star+116) \right. \nonumber \\
   && +\left.3 x^2 (-245 W_p+1365 W_\star+72)-210 W_\star+48\right) \nonumber \\
   && +r^6 \left(-5 x^6 \left(2989 W_p^2-2 W_p (5929 W_\star+928)+W_\star (2989 W_\star+928)\right) \right. \nonumber \\
   && +2 x^4 \left(7245 W_p^2+W_p (1240-10710 W_\star)-48 W_\star (105 W_\star+59)\right) \nonumber \\
   && +3 x^2 \left(21 W_p^2-14 W_p (103 W_\star+24)+W_\star (2681 W_\star+416)-128\right) \nonumber \\
   && +\left. 17640 W_p x^8 (W_\star-2 W_p)+6 (16-35 W_\star) W_\star+48\right) \nonumber \\
   && +2 r^5 x \left(2 x^2 \left(-1512 W_p^2+21 W_p (271 W_\star+17)+W_\star (546 W_\star+995)+384\right)\right. \nonumber \\
   && +8820 W_p^2 x^8+x^4 \left(-21735 W_p W_\star-4 W_p (1645 W_p+2468)+9730 W_\star^2+7688 W_\star\right) \nonumber \\
   && +\left.40 W_p x^6 (784 W_p-637 W_\star-116)+3 (W_p (91 W_\star+54)-6 W_\star (91 W_\star+29)+24)\right) \nonumber \\
   && +r^4 \left(-42140 W_p^2 x^8+x^4 \left(4 (2429 W_p-4894) W_\star+16 (W_p (966 W_p+841)-96)-12145 W_\star^2\right)\right. \nonumber \\
   && +2 W_p x^6 (-19390 W_p+28385 W_\star+13056)+14 W_\star x^2(-589 W_p+65 W_\star-10) \nonumber \\
   && +\left.8 (W_p (98 W_p-221)-240) x^2+3 W_\star (161 W_\star+76)+40\right) \nonumber \\
   && +2 r^3 x \left(x^2 \left(-2513 W_p^2+14 W_p (61 W_\star-98)+10 W_\star (182 W_\star+597)+1440\right)+20055 W_p^2 x^6 \right. \nonumber \\
   && +\left.W_p x^4 (4018 W_p-15365 W_\star-14208)+511 W_p W_\star+180 W_p-196 W_\star^2-314 W_\star+352\right) \nonumber \\
   && +r^2 \left(-x^2 \left(-539 W_p^2+8 W_p (105 W_\star+46)+12 W_\star(35 W_\star+291)+1952\right) \right. \nonumber \\
   && -\left. 19005 W_p^2 x^6+2 W_p x^4 (413 W_p+4060 W_\star+7520)+4 W_\star (7 W_\star+33)-64\right) \nonumber \\
   && +4 r x \left(W_p \left(-x^2 (112 W_p+210 W_\star+971)+1120 W_p x^4+14 W_\star+33\right)+98 W_\star+140\right) \nonumber \\
   && +\left.\left.28 W_p x^2 \left(-15 W_p x^2+W_p+14\right)-56\right)\right] 
   \end{eqnarray}
   \begin{eqnarray}
A_{33}&=&\frac{1}{112 \left(r^2-2 r x+1\right)^2} 
   \left[14 r^8 x^2 \left(21 x^2 \left(11 x^4-15 x^2+5\right)-5\right) (W_p-W_\star)^2 \right. \nonumber \\
   && +2 r^7 x (W_p-W_\star) \left(1617 x^8 (W_\star-5 W_p)+21 x^6 (399 W_p+175 W_\star+52)-5 x^4 (147 W_p+1323 W_\star+176)\right. \nonumber \\
   && +\left. x^2 (-455 W_p+2065 W_\star-12)-70 W_\star+24\right) \nonumber \\
   && +r^6 \left(-21 x^8 \left(119 W_p^2+2 W_p (945 W_\star+208)-W_\star (553 W_\star+208)\right) \right. \nonumber \\
   && +x^6 \left(-23275 W_p^2+W_p (47726 W_\star+2248)+W_\star (2303 W_\star+3456)\right) \nonumber \\
   && +x^4 \left(6055 W_p^2+W_p (3664-2030 W_\star)-21 W_\star (695 W_\star+256)+704\right) \nonumber \\
   && +x^2 \left(119 W_p^2-6 W_p (413 W_\star+52)+W_\star (4109 W_\star+192)-528\right) \nonumber \\
   && -\left. 12936 W_p x^{10} (W_\star-2 W_p)-70 W_\star^2+48 W_\star+48\right) \nonumber \\
   && +2 r^5 x \left(x^6 \left(20776 W_p^2+W_p (8379 W_\star+8248)-2 W_\star (4067 W_\star+4034)\right)\right. \nonumber \\
   && +x^4 \left(2660 W_p^2-W_p (24745 W_\star+7297)+W_\star (3094 W_\star+2165)-1408\right) \nonumber \\
   && -6468 W_p^2 x^{10}+x^2 \left(4025 W_p W_\star-2 W_p (742 W_p+333)+3654 W_\star^2+2678 W_\star+616\right) \nonumber \\
   && -\left.84 W_p x^8 (259 W_p-238 W_\star-52)+189 W_p W_\star+51 W_p-854 W_\star^2-135 W_\star+120\right) \nonumber \\
   && +r^4 \left(x^6 \left(-39130 W_p^2+2 W_p (7329 W_\star-724)+W_\star (11123 W_\star+23148)+2816\right) \right. \nonumber \\
   && +x^4 \left(3612 W_p^2+6 W_p (3591 W_\star+2308)-W_\star (6307 W_\star+12354)+2816\right) \nonumber \\
   && -x^2 \left(-434 W_p^2+2 W_p (1785 W_\star+88)+W_\star (1715 W_\star+1888)+2312\right) \nonumber \\
   && +\left.33516 W_p^2 x^{10}+2 W_p x^8 (8624 W_p-24157 W_\star-13952)+259 W_\star^2+54 W_\star+40\right) \nonumber \\
   && +2 r^3 x \left(x^4 \left(7455 W_p^2-W_p (7749 W_\star+7178)-W_\star (1862 W_\star+8065)-3168\right) \right. \nonumber \\
   && +x^2 \left(-917 W_p^2-W_p (2149 W_\star+2276)+14 W_\star (78 W_\star+331)+480\right) \nonumber \\
   && -\left. 17395 W_p^2 x^8+W_p x^6 (4585 W_p+14357 W_\star+17264)+245 W_p W_\star+30 W_p+98 W_\star^2+71W_\star+448\right) \nonumber \\
   && +r^2 \left(x^4 \left(-2597 W_p^2+8 W_p (609 W_\star+1382)+490 W_\star^2+5478 W_\star+5344\right)\right. \nonumber \\
   && +x^2 \left(231 W_p^2+W_p (420 W_\star+416)-12 (W_\star (21 W_\star+233)+160)\right) \nonumber \\
   && +\left. 18081 W_p^2 x^8-W_p x^6 (9443 W_p+8428 W_\star+20880)+2 (3-7 W_\star) W_\star-64\right) \nonumber \\
   && +2 r x \left(2 x^2 \left(56 W_p^2-14 (9 W_p+13) W_\star-769 W_p-500\right)\right. \nonumber \\
   && -\left. 2352 W_p^2 x^6+W_p x^4 (1344 W_p+490 W_\star+3103)-14 W_p W_\star+3 W_p+140 W_\star+328\right) \nonumber \\
   && +\left. 14 x^2 \left(W_p \left(35 W_p x^4-2 (9 W_p+26) x^2-W_p+20\right)+20\right)-56\right]
\end{eqnarray}
\begin{eqnarray}
A_{04}&=&\frac{35 r^4 \left(x^2-1\right)^4 (r (W_p x (r-2 x)-r W_\star x+W_\star)+W_p x)^2}{256 \left(r^2-2 r x+1\right)^2} \\
A_{14}&=&-\frac{1}{64 \left(r^2-2 r x+1\right)^2} \nonumber \\
   && \left[5 r^2 \left(x^2-1\right)^3 (r (W_p x (r-2 x)-r W_\star x+W_\star)+W_p x) \left(7 r^4 x \left(9 x^2-1\right) (W_p-W_\star)\right.\right. \nonumber \\
   && +r^3 \left(-42 W_p \left(3 x^4+x^2\right)+7 W_\star \left(17 x^2-1\right)+4\right) \nonumber \\
   && +\left.\left.r^2 x \left(5 W_p \left(35 x^2+1\right)-68 W_\star-8\right)+4 r \left(-20 W_p x^2+3 W_\star+1\right)+12 W_p x\right)\right] \\
A_{24}&=&  \frac{1}{128 \left(r^2-2 r x+1\right)^2} 
   \left[3 \left(x^2-1\right)^2 \left(35 r^8 x^2 \left(33 x^4-18 x^2+1\right) (W_p-W_\star)^2 \right.\right. \nonumber \\
   && +10 r^7 x (W_p-W_\star) \left(21 x^4 (4 W_p+19 W_\star)+14 x^2 (3 W_p-13 W_\star+2)-462 W_p x^6+7 W_\star-4\right) \nonumber \\
   && +r^6 \left(5 \left(924 W_p^2 x^8+7 x^4 \left(4 (7 W_p+4) W_\star-8 W_p (9 W_p+4)+153 W_\star^2\right) \right.\right. \nonumber \\
   && +42 W_p x^6 (31 W_p-65 W_\star)+2 W_\star x^2 (115 W_p-187 W_\star+44) \nonumber \\
   && +\left.\left.W_\star (7 W_\star-8)\right)-10 W_p (5 W_p+8) x^2+8\right) \nonumber \\
   && +2 r^5 x \left(10 x^2 \left(65 W_p^2+W_p (68-155 W_\star)-22 W_\star (8 W_\star+3)\right)\right. \nonumber \\
   && -\left.5670 W_p^2 x^6+35 W_p x^4 (-28 W_p+225 W_\star+16)-85 W_p W_\star-16 W_p+400 W_\star^2-84 W_\star-16\right) \nonumber \\
   && +r^4 \left(x^2 \left(-197 W_p^2+32 W_p (47 W_\star-16)+8 W_\star (146 W_\star+137)+32\right)\right. \nonumber \\
   && +\left. 11235 W_p^2 x^6-10 W_p x^4 (103 W_p+904 W_\star+208)+8 \left(-15 W_\star^2+W_\star+2\right)\right) \nonumber \\
   && +8 r^3 x \left(-710 W_p^2 x^4+W_p x^2 (86 W_p+334 W_\star+179)-28 W_p W_\star+7 W_p-22 W_\star^2-48 W_\star-4\right) \nonumber \\
   && +8 r^2 \left(189 W_p^2 x^4-W_p x^2 (13 W_p+46 W_\star+54)+W_\star^2+6 W_\star+1\right) \nonumber \\
   && +\left.\left.16 r W_p x \left(-12 W_p x^2+W_\star+3\right)+8 W_p^2 x^2\right)\right]
\end{eqnarray}
\begin{eqnarray}
A_{34}&=&-\frac{1}{64 \left(r^2-2 r x+1\right)^2} 
   \left[\left(x^2-1\right) \left(7 r^8 x^2 \left(429 x^6-495 x^4+135 x^2-5\right) (W_p-W_\star)^2 \right.\right. \nonumber \\
   && +2 r^7 x (W_p-W_\star) \left(231 x^6 (18 W_p+25 W_\star)+315 x^4 (2 W_p-19 W_\star+2)-35 x^2 (10 W_p-39 W_\star+12) \right. \nonumber \\
   && -\left. 6006 W_p x^8-35 W_\star+30\right)+r^6 \left(35 x^4 \left(74 W_p^2+6 W_p(29 W_\star+8)+9 (4-51 W_\star) W_\star\right) \right. \nonumber \\
   && +5 x^2 \left(22 W_p^2+W_p (96-366 W_\star)+3 W_\star (187 W_\star-96)+24\right)+12012 W_p^2 x^{10} \nonumber \\
   && +21 x^6 \left(6 (177 W_p+20) W_\star-30 W_p (31 W_p+8)+851 W_\star^2\right)+462 W_p x^8 (31 W_p-87 W_\star) \nonumber \\
   && -\left.35 W_\star^2+60 W_\star-24\right)+2 r^5 x \left(7 x^4 \left(850 W_p^2+5 W_p (84-535 W_\star) \right.\right. \nonumber \\
   && -\left. 2 W_\star (508 W_\star+255)\right)-17094 W_p^2 x^8 \nonumber \\
   && +5 x^2 \left(3 (55 W_p+52) W_\star-2 W_p (89 W_p+180)+1056 W_\star^2-48\right) \nonumber \\
   && +\left. 21 W_p x^6 (186 W_p+1295 W_\star+120)+145 W_p W_\star-12 W_p-600 W_\star^2+342 W_\star\right) \nonumber \\
   && +r^4 \left(5 x^4 \left(-363 W_p^2+4928 W_p W_\star+4 W_\star (307 W_\star+387)+96\right) +40299 W_p^2 x^8\right. \nonumber \\
   && -7 W_p x^6 (3095 W_p+5464 W_\star+1680)-24 W_\star x^2 (73 W_p+146 W_\star+103) \nonumber \\
   && +(7 W_p (43 W_p+240)+528) x^2 \nonumber \\
   && +\left.12 W_\star (15 W_\star-7)-32\right)+4 r^3 x \left(-2 x^2 \left(63 W_p^2+W_p (972 W_\star+309)+2 W_\star (85 W_\star+252)+108\right)\right. \nonumber \\
   && -\left. 6286 W_p^2 x^6+5 W_p x^4 (716 W_p+738 W_\star+543)+78 W_p W_\star-57 W_p+132 W_\star^2+264 W_\star-40\right) \nonumber \\
   && +4 r^2 \left(x^2 \left(33 W_p^2+276 W_p (W_\star+1)+6 W_\star (5 W_\star+42)+142\right) \right. \nonumber \\
   && +\left.2185 W_p^2 x^6-2 W_p x^4 (537 W_p+370 W_\star+618)-6 W_\star (W_\star+6)+2\right) \nonumber \\
   && +16 r x \left(W_p \left(3 x^2 (12 W_p+5 W_\star+23)-100 W_p x^4-3(W_\star+3)\right)-2 (3 W_\star+5)\right) \nonumber \\
   && +\left.\left.24 W_p x^2 \left(W_p \left(5 x^2-1\right)-4\right)+16\right)\right]
\end{eqnarray}
\begin{eqnarray}
A_{44}&=&\frac{1}{256 \left(r^2-2 x r+1\right)^2} 
   \left[(W_p-W_\star)^2 x^2 \left(6435 x^8-12012 x^6+6930 x^4-1260 x^2+35\right) \right. \nonumber \\
   &&  r^8+2 (W_p-W_\star) x \left(-12870 W_p x^{10}+429 (40 W_p+31 W_\star) x^8-924 (3 W_p+25 W_\star-2) x^6 \right. \nonumber \\
   && -\left. 630 (4 W_p-19 W_\star+4) x^4+70 (7 W_p-26 W_\star+12) x^2+5 (7 W_\star-8)\right) r^7 \nonumber \\
   && +\left(25740 W_p^2 x^{12}+858 W_p (23 W_p-109 W_\star) x^{10} \right. \nonumber \\
   && +33 \left(1363 W_\star^2+8 (421 W_p+28) W_\star-448 W_p (5 W_p+1)\right) x^8 \nonumber \\
   && +84 \left(405 W_p^2+(168-87 W_\star) W_p-W_\star (851 W_\star+4)\right) x^6 \nonumber \\
   && -70 \left(26 W_p^2+260 W_\star W_p+3 (40-153 W_\star) W_\star-8\right) x^4 \nonumber \\
   && -\left.10 \left(374 W_\star^2-251 W_p W_\star-296 W_\star+W_p (17 W_p+112)+48\right) x^2 +35 W_\star^2 -80 W_\star+48\right) r^6 \nonumber \\
   && +2 x \left(-40326 W_p^2 x^{10}+33 W_p (984 W_p+2115 W_\star+224) x^8 \right. \nonumber \\
   && +12 \left(1365 W_p^2+(476-7805 W_\star) W_p-2 W_\star (832 W_\star+483)\right) x^6 \nonumber \\
   && -14 \left(940 W_p^2+(960-1975 W_\star) W_p-4 W_\star (508 W_\star+171)+80\right) x^4 \nonumber \\
   && +10 \left(105 W_p^2+(90 W_\star+296) W_p+12 (7-88 W_\star) W_\star+64\right) x^2 \nonumber \\
   && +\left.800 W_\star^2+64 W_p-205 W_p W_\star-744 W_\star+96\right) r^5 \nonumber \\
   && +\left(105699 W_p^2 x^{10}-12 W_p (10241 W_p+9176 W_\star+3248) x^8+14 \left(1415 W_p^2+96 (107 W_\star+16) W_p \right.\right. \nonumber \\
   && +\left.8 W_\star (176 W_\star+255)+160\right) x^6+20 \left(337 W_p^2+8 (74-277 W_\star) W_p-4 W_\star (307 W_\star+337)+88\right) x^4 \nonumber \\
   && +\left.\left(-397 W_p^2+1664 (W_\star-2) W_p+48 (W_\star (146 W_\star+69)-44)\right) x^2-16 (W_\star-1) (15 W_\star+2)\right) r^4 \nonumber \\
   && +16 x \left(-4638 W_p^2 x^8+7 W_p (806 W_p+434 W_\star+369) x^6 \right. \nonumber \\
   && -\left(322 W_\star^2+4 (905 W_p+272) W_\star+5 W_p (290 W_p+427)+300\right) x^4 \nonumber \\
   && +\left(-2 W_p^2+(942 W_\star+81) W_p+4 W_\star (85 W_\star+242)+88\right) x^2 \nonumber \\
   && -\left. 66 W_\star^2+31 W_p-24 W_p W_\star-120 W_\star+52\right) r^3 \nonumber \\
   && +16 \left(1841 W_p^2 x^8-W_p (2115 W_p+714 W_\star+1378) x^6 \right. \nonumber \\
   && +\left(507 W_p^2+4 (185 W_\star+289) W_p+5 W_\star (7 W_\star+66)+243\right) x^4 \nonumber \\
   && -\left. (3 W_p (3 W_p+38)+118) x^2-6 W_\star (23 W_p+5 W_\star+42) x^2+3 W_\star (W_\star+6)-5\right) r^2 \nonumber \\
   && +32 x \left(-196 W_p^2 x^6+5 W_p (40 W_p+7 W_\star+37) x^4-2 (10 W_\star+3 W_p (6 W_p+5 W_\star+23)+22) x^2 \right. \nonumber \\
   && +\left.\left.9 W_p+3 W_p W_\star+12 W_\star+20\right) r+16 \left(35 W_p^2 x^6-10 W_p (3 W_p+4) x^4+3 (W_p (W_p+8)+4)x^2-4\right)\right] \nonumber \\ 
\end{eqnarray}
and 
\begin{eqnarray}
B_{00}&=&\frac{1}{7 \left(r^2-2 r x+1\right)}\left[2 r x^3 \left(7 W_p\left(-2 r^2+W-2\right)+4 r^2 W_\star\right)\right. \nonumber \\
 &&+ x^2 \left(7 \left(r^2+1\right) W_p\left(2r^2-W+2\right)-2 r^2 \left(7 r^2+11\right) W_\star\right) \nonumber \\
 &&+ \left.2 r \left(17 r^2+7\right) W_\star x-20 r^2 W_\star\right] \\
B_{01}&=&\frac{1}{7 \left(r^2-2 r x+1\right)^2 \left(r^2+2 r x+1\right)} \left[\left(x^2-1\right) \left(4 r^2 \left(r^2+1\right) x^4 
      \left(W_p\left(11 r^2-7 W+7\right)-11 r^2W_\star\right)\right.\right. \nonumber \\
   && +r^2 \left(r^2+1\right) \left(23 r^2+17\right) W_\star+8 r^3 x^5 \left(W_p\left(-11 r^2+7W-7\right)+2 r^2 W_\star\right) \nonumber \\
   && +x^2 \left(7 \left(r^2+1\right)^3 W_p^2-\left(14 r^6+23 r^4+28 r^2+7\right)\left(r^2+1\right) W_p 
      +r^2 \left(14 r^6+21 r^4-38 r^2+11\right) W_\star\right)\nonumber \\
   && +2 r x^3 \left(\left(r^4+55r^2+12\right) r^2 W_\star+W_p\left(14r^6+23r^4-7\left(r^2+1\right)^2 W_p+28 r^2+7\right)\right) \nonumber \\
   && \left.\left.-r \left(37r^6+63r^4+45r^2+7\right) W_\star x\right)\right]
\end{eqnarray}
\begin{eqnarray}
B_{11}&=&\frac{3}{7 \left(r^2-2 r x+1\right)^2\left(r^2+2 r x+1\right)}
    \left[4 r^2 \left(r^2+1\right) x^6 \left(W_p\left(-11 r^2+7 W-7\right)+11 r^2 W_\star\right) \right. \nonumber \\
   && +x^2 \left(\left(r^2+1\right) \left(19 r^4+14 r^2+7\right) W-4 r^2 \left(13 r^4+14 r^2+7\right) W_\star\right) \nonumber \\
   && +3 r^2 \left(r^4-1\right) W_\star+8 r^3 x^7 \left(W_p\left(11 r^2-7 W+7\right)-2 r^2 W_\star\right) \nonumber \\
   && -2 r x^5 \left(\left(r^4+63 r^2+12\right) r^2 W_\star+W_p\left(14 r^6+11 r^4-7 \left(r^2+1\right)^2 W+7\right)\right) \nonumber \\
   && +x^4\left(\left(r^2+1\right) W_p\left(14 r^6+11 r^4-7 \left(r^2+1\right)^2 W+7\right)-r^2 \left(14 r^6+9 r^4-82
   r^2+11\right) W_\star\right) \nonumber \\
   && +r x^3 \left(\left(47 r^6+53 r^4+21 r^2+7\right) W_\star-2 \left(19 r^4+14 r^2+7\right) W\right) \nonumber \\
   && +\left.r \left(-3 r^6+19 r^4+17 r^2+7\right) W_\star x\right] \\
B_{02}&=&\frac{3}{56 \left(r^2-2 r x+1\right)^2 \left(r^2+2 r x+1\right)} 
    \left[\left(x^2-1\right)^2 \left(-2 r^2 \left(r^2+1\right) \left(13 r^2+7\right) W_\star \right.\right. \nonumber \\
   && +4 r^2 x^4 \left(8 r^4 W_\star-\left(r^2+1\right) W_p\left(8 r^2-7 W\right)\right) \nonumber \\
   && -2 r x^3 \left(W_p\left(2 r^2 \left(7 r^4+2 r^2+7\right)-7 \left(r^2+1\right)^2 W_p\right)+2 \left(3 r^2+25\right) r^4 W_\star\right) \nonumber \\
   && +8 r^3 W_px^5 \left(8 r^2-7 W\right)+x^2 \left(2 \left(-7 r^4+4 r^2+27\right) r^4 W_\star \right. \nonumber \\
   && +\left.\left.\left.W_p\left(2 r^2 \left(7 r^6+9 r^4+9 r^2+7\right)-7
   \left(r^2+1\right)^3 W_p\right)\right)+4 r^3 \left(10 r^4+11 r^2+7\right) W_\star x\right)\right] \\
B_{12}&=&\frac{1}{28 \left(r^2-2 r x+1\right)^2 \left(r^2+2 r x+1\right)} 
   \left[\left(x^2-1\right) \left(6 r^2 \left(r^2+1\right) \left(r^2+11\right) W_\star \right.\right. \nonumber \\
   && +60 r^2 x^6 \left(\left(r^2+1\right) W_p\left(8 r^2-7 W_p\right)-8 r^4 W_\star\right) 
      +120 r^3 W_px^7 \left(7 W_p-8 r^2\right)+2 rx^5 \left(30 r^6 (7 W_p+3 W_\star) \right. \nonumber \\
   && +\left.3 r^4 (314 W_\star-5 W_p(7 W_p+44))-2 r^2 W_p(91 W_p+167)-105 W_p^2\right) \nonumber \\
   && +2 r x^3 \left(-7 \left(r^2+1\right)^2 W_p^2-2 \left(201 r^4-42 r^2+5\right) r^2 W_\star+2 \left(63 r^6+290 r^4+223
   r^2+56\right) W_p\right) \nonumber \\
   && +x^2 \left(7 \left(r^2+1\right)^3 W_p^2-2 \left(63 r^6+290 r^4+223 r^2+56\right) \left(r^2+1\right) W_p \right. \nonumber \\
   && +\left.2 r^2 \left(63 r^6+587 r^4+469 r^2+233\right) W_\star\right)+x^4 \left(2 r^4 \left(105 r^2 \left(r^2-4\right)-869\right) W_\star \right. \nonumber \\  
   && \left.-\left(r^2+1\right) W_p\left(210 r^6-15 r^4 (7 W_p+44)-2 r^2 (91 W_p+167)-105 W_p\right)\right) \nonumber \\
   && -4 r \left.\left.\left(33 r^6+163 r^4+128 r^2+28\right) W_\star x\right)\right] \\
B_{22}&=&\frac{1}{56} \left[\frac{1}{r^4+r^2 \left(2-4 x^2\right)+1}\right. \nonumber \\
   && \times \left(2 W_px^2 \left(7 r^6 \left(35 x^4+10 x^2-21\right)-2 r^4 \left(280 x^6+405 x^4-866 x^2+265\right) \right.\right. \nonumber \\
   && +r^2 \left.\left(-747 x^4+934 x^2-355\right)+56 \left(4 x^2-1\right)\right) \nonumber \\
   && -\frac{1}{\left(r^2-2 r x+1\right)^2} \left(2 r W_\star (r x-1) \left(7 r^4 x \left(35 x^4+10 x^2-21\right)+r^3 \left(-280 x^6-1055 x^4+618 x^2+45\right)\right.\right. \nonumber \\
   && +\left.\left.r^2 x \left(1055 x^4+426 x^2-473\right)+r \left(-901 x^4+142 x^2+87\right)+56 x \left(4 x^2-1\right)\right)\right) \nonumber \\
   && +\left.7 W_p^2 \left(-35 x^4+6 x^2+5\right) x^2\right] \\
B_{13}&=&-\frac{3}{8 \left(r^2-2 r x+1\right)} 
  \left[\left(x^2-1\right)^2 \left(-14 r^3 W_\star x+2 r W_px^3 \left(10 r^2-W\right)+4 r^2 W_\star\right.\right. \nonumber \\
  && +\left.\left. x^2 \left(10 r^4 (W_\star-W)+r^2 (W-10) W+W_p^2\right)\right)\right] \\
B_{23}&=&-\frac{1}{4 \left(r^2-2 r x+1\right)} 
  \left[-3 \left(5 x^4-6 x^2+1\right) \left(W_p^2 x^2 \left(r^2-2 r x+1\right)-4 r^2 W_\star (r x-1)\right)\right. \nonumber \\
  && +10 r^2 x\left(7 x^4-10 x^2+3\right) (r (W_px (r-2 x)-r W_\star x+W_\star)+W_px) \nonumber \\
  && -\left. 12 x \left(1-x^2\right) (r (W_px (r-2 x)-r W_\star x+W_\star)+W_px)\right] \\
B_{33}&=&-\frac{1}{8 \left(r^2-2 r x+1\right)} 
   \left[\left(35 x^4-30 x^2+3\right) \left(W_p^2 x^2 \left(r^2-2 r x+1\right)-4 r^2 W_\star (r x-1)\right)\right. \nonumber \\
  && -2 r^2 x \left(63 x^4-70 x^2+15\right) (r (W_px (r-2 x)-r W_\star x+W_\star)+W_px) \nonumber \\
  && +\left. 8 x \left(3-5 x^2\right) (r (W_px (r-2 x)-r W_\star x+W_\star)+W_px)\right]
\end{eqnarray}
\end{widetext}
\bibliographystyle{apsrev}
\bibliography{ref}

\end{document}